\documentclass[apj]{emulateapj}
\usepackage{lscape}
\def\gsim{\ \raise 3pt \hbox{$\rangle$} \kern -8.5pt \raise -2pt \hbox{$\sim$}\ }
\usepackage{graphicx,natbib,color,amsmath,subfigure}
\newcommand{\tw}{\bf \color{red}}
\newcommand{\au}{_{\rm au}}
\newcommand{\sfu}{\rm sfu}

\begin{document}
\title{Sub-THz radiation mechanisms in solar flares}
\author{Gregory D. Fleishman\altaffilmark{1,2} and Eduard P. Kontar\altaffilmark{3}}
\altaffiltext{1}{Center for Solar-Terrestrial Research, New Jersey
    Institute of Technology, Newark, NJ 07102; gfleishm@njit.edu}
\altaffiltext{2}{Ioffe Physical-Technical Institute of the Russian
Academy of Sciences, St. Petersburg 194021, Russia} 
\altaffiltext{3}{Department of Physics and Astronomy, University of
Glasgow, G12 8QQ, United Kingdom}

\begin{abstract}

Observations in the sub-THz range of large solar flares have revealed a
mysterious spectral component increasing with frequency and hence
distinct from the microwave component commonly accepted to be
produced by gyrosynchrotron (GS) emission from accelerated electrons.
Evidently, having a distinct sub-THz component requires
either a distinct emission mechanism (compared to the GS one), or
different properties of electrons and location, or both. We
find, however, that the list of possible emission mechanisms is
incomplete. This Letter proposes a more complete list of emission
mechanisms, capable of producing a sub-THz component, both
well-known and new in
this context and  calculates a representative set of their spectra 
produced by a) free-free emission, b) gyrosynchrotron
emission, c) synchrotron emission from relativistic
positrons/electrons, d) diffusive radiation, and e) Cherenkov
emission. We discuss the possible role
of the mechanisms in forming the sub-THz emission and emphasize
their diagnostics potential for flares.

\end{abstract}

\keywords{Sun: flares---acceleration of
particles---turbulence---diffusion---Sun: magnetic fields---Sun:
radio radiation}

\section{Introduction}


Solar flares, being manifestations of prompt energy releases,
produce electromagnetic radiation throughout the entire spectrum of
electromagnetic emission from radio waves to gamma rays
\citep[e.g.][]{DennisSchwartz1989,BrownKontar2005}. Although the
radio, UV, X-ray, and gamma-ray emission is generally well observed
with high resolution from huge number of events, the sub-THz
radiation has only recently been observed from a few large events,
at a small number of frequencies
\citep{Kaufmann_etal2001,Trottet_etal_2002,Luethi_etal_2004a,
Luethi_etal_2004b,Kaufmann_etal_2004,Cristiani_etal_2008,Trottet_etal_2008,
Kaufmann_etal_2009a,Kaufmann_etal_2009fast,Silva_etal_2007}. 
The available observational tools are unable to measure polarization and
are clearly insufficient to provide detailed spectral and positional
information about the sub-THz bursts. Nevertheless, available
sub-THz observations have already provided us with puzzling
questions which have yet to be answered.

The observations suggest that, on top of quiet Sun 
emission, at least two kinds of sub-THz emission can be
produced. The first kind looks like a natural extension of the
microwave spectrum at higher frequencies and so can reasonably be
interpreted as synchrotron radiation from accelerated electrons,
which are also responsible for microwave and hard X-ray emission.
The second kind 
looks like a distinct spectral component rising with
frequency in the sub-THz range in contrast to the microwave spectrum,
which falls with frequency. The origin of this component is unclear;
there is no consensus about the possible emission mechanism producing it,
moreover, it is ambiguous whether the emission is of thermal
or nonthermal origin.

The main observational characteristics of this component are: relatively
large radiation peak flux  of the order of 10$^4$~sfu
\citep{Kaufmann_etal_2004};  radiation spectrum  rising with
frequency $F(f){\propto} f^\delta$;  spectral index  varying with
time within $\delta\sim1\dots6$;  sub-THz component can display a
sub-second time variability with the modulation about $5\%$
\citep{Kaufmann_etal_2009fast}; the source size is believed to be
less than $20''$, however, this conclusion is based on a multi-beam
observation of a few antennas with $\sim4'$ resolution each, rather
than on true imaging; therefore, the size estimate must be
considered with caution. In addition, observations \citep{Luethi_etal_2004a,Luethi_etal_2004b}
suggest the existence of both compact $\sim 10''$
and extended $\sim 60''$ components, and source sizes increasing
with frequency.

The emission mechanisms proposed so far to account for the sub-THz
component are thermal free-free emission, GS emission from
flare accelerated electrons, and synchrotron emission from
nuclear decay-generated relativistic positrons \citep[e.g.][as a
review]{Nindos_etal_2008}. None of these mechanisms is readily
consistent with the full list of the sub-THz component properties
and/or with available context observations at other wavelength.
Below we analyze these options 
and also consider two other emission mechanisms capable of
producing a spectrum rising with frequency---diffusive radiation in
Langmuir (DRL) waves \citep[e.g.][]{Fl_Topt_2007_MNRAS,
Fl_Topt_2007_PhRvE} and Vavilov-Cherenkov emission
\citep[e.g.,][]{Bazylev_Zhevago_1987} from  chromospheric layers,
which make the list of options more complete. Calculating the
spectrum from all of the above mentioned models, we establish the range of
main source parameters for each model emphasizing the strengths and
weaknesses of each emission mechanism.

\begin{figure*}[t]
\begin{center}
\epsscale{0.9} \plottwo{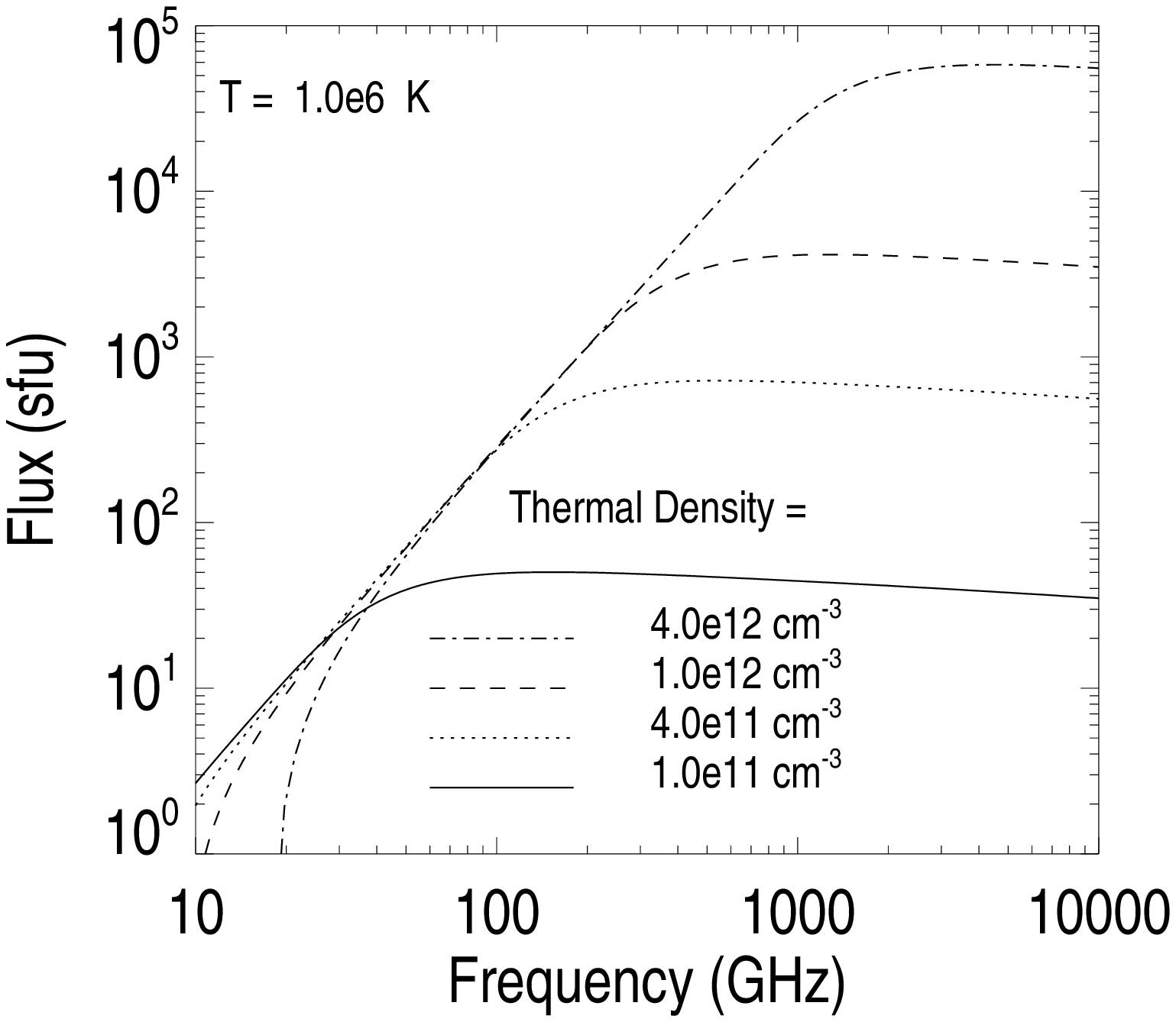}{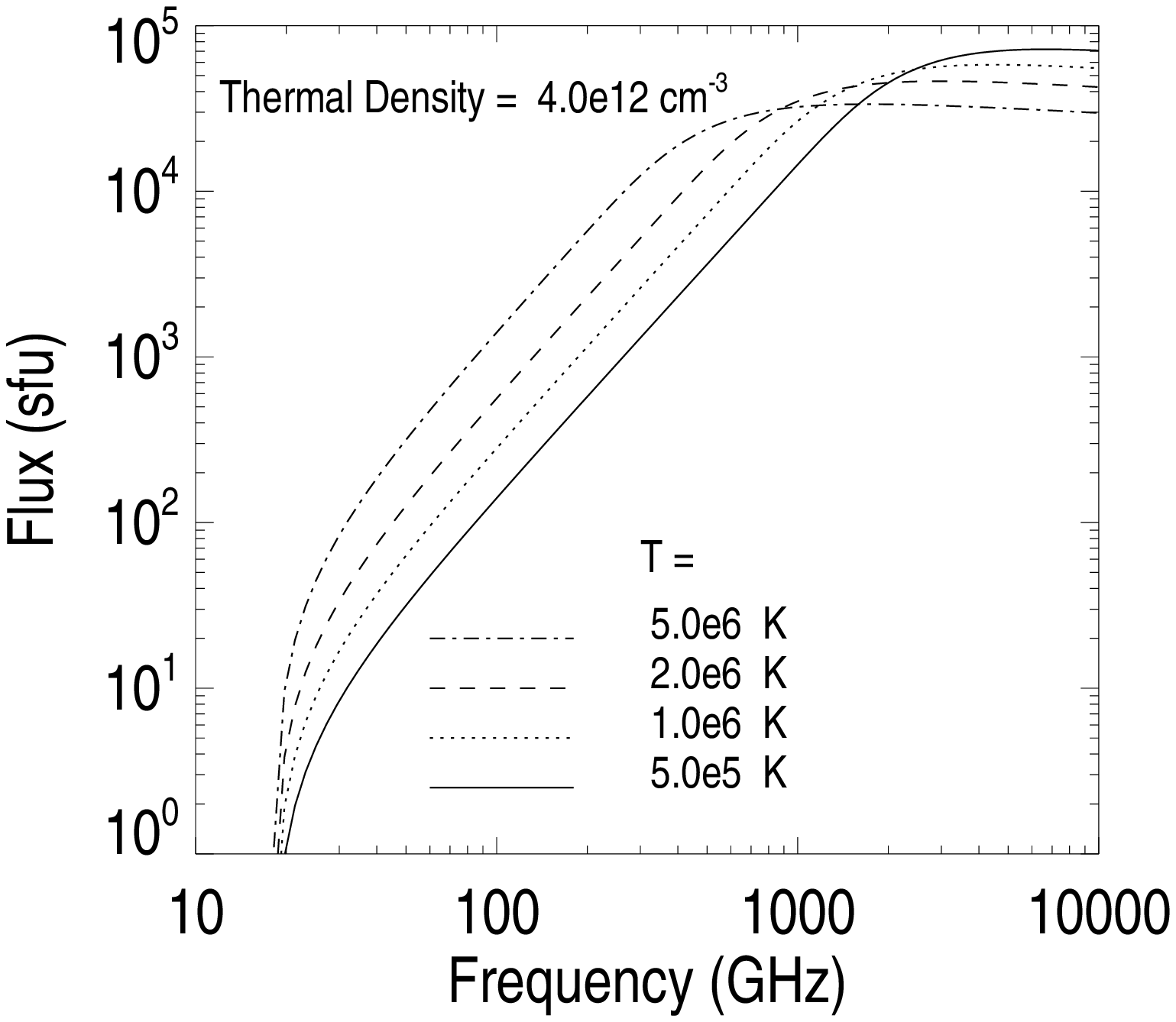} \caption{Radio spectra
produced by thermal free-free emission from a uniform cubic source
with a linear size of $20''$ for
$n_e=10^{11}\dots4\cdot10^{12}$~cm$^{-3}$ and $T_e=0.5\dots 5$~MK. }
\label{FIG_ff}
\end{center}
\end{figure*}

\section{Free-Free Emission}

Perhaps the simplest example of a radio spectrum rising with frequency
is optically thick thermal free-free emission. Having a rising
spectrum from a compact ($\lesssim20''$) source requires obviously
that the source is relatively dense ($n_e{\gtrsim}
10^{11}$~cm$^{-3}$) and hot ($T_e{\gtrsim}10$~MK). 
From available X-ray observations
\citep[e.g.][]{DennisSchwartz1989,Kasparova_etal2005}, we can
exclude the option of a source that is simultaneously dense and hot,
say $n_e\sim10^{12}$~cm$^{-3}$ and $T_e\sim10$~MK,
$EM=n_e^2V\sim3\times10^{51}$~cm$^{-3}$, since a plasma with the
corresponding emission measure and temperature would produce
stronger X-ray emission than is actually observed from solar flares
\citep{DennisSchwartz1989,Emslie_etal2003,Brown_etal2007}. We cannot
exclude, however, dense plasmas of lower temperature, $T_e\sim1$~MK;
such plasmas do not contradict to the current X-ray and UV observations.

Figure~\ref{FIG_ff} shows free-free spectra from overdense 
sources with temperature around 1~MK. Evidently, having a flux
density above 1000~sfu level requires  thermal electron number
density above $10^{12}$~cm$^{-3}$ or/and the linear size of the
source above $20''$. Although available size estimates do not favor
sources larger than $20''$, a firm conclusion about that must
await true imaging observations at the sub-THz range. Such large
emission sources, tens of arcseconds, are consistent with at least
some observations \citep{Luethi_etal_2004a,Luethi_etal_2004b}.


It should be noted that the spectral index of a spatially uniform
source of free-free radiation cannot be larger than $2$, while the
observations suggest larger values, especially during the initial phase
of the sub-THz bursts \citep{Kaufmann_etal_2009a}. These larger
values can be reconciled with the free-free mechanism if we allow
for a cooler dense absorbing layer in the line of sight between the
source and observer, providing an attenuation factor
$\exp(-\tau_{ff})$, where
$\tau_{ff}{\propto}nn_e(1-\exp(-hf/kT))T^{-1/2}f^{-3}\propto{nn_e}{T^{-3/2}f^{-2}}$
is the free-free optical depth; such a non-uniform source is capable
of producing  frequency spectra with a very sharp low frequency edge
compatible with observations.

Finally, we emphasize that the free-free emission mechanism can
display temporal variability. 
For sausage mode loop oscillations, for example,
$B=B_0(1+m\cos(2{\pi}t/P))$, with a period $P$ and modulation
amplitude $m<1$, we can calculate all relevant parameter variations
and, thus, make firm prediction about the free-free emission
modulation amplitude in the optically thick and thin regimes:
$F_{thick}\propto(1+\frac{m}{6}\cos(2{\pi}t/P))$ and
$F_{thin}\propto(1+\frac{2m}{3}\cos(2{\pi}t/P))$. Therefore, the
optically thick and thin part of the free-free emission oscillate in
phase. However, the modulation amplitude in the thin regime is four
times larger than in the thick regime.

%

We can conclude that the free-free model can provide sufficient flux
and spectral index consistent with observations of the sub-THz
component, although the source density must be somewhat high
compared with standard coronal loop densities. 

\begin{figure*}[t]
\begin{center}
 \plottwo{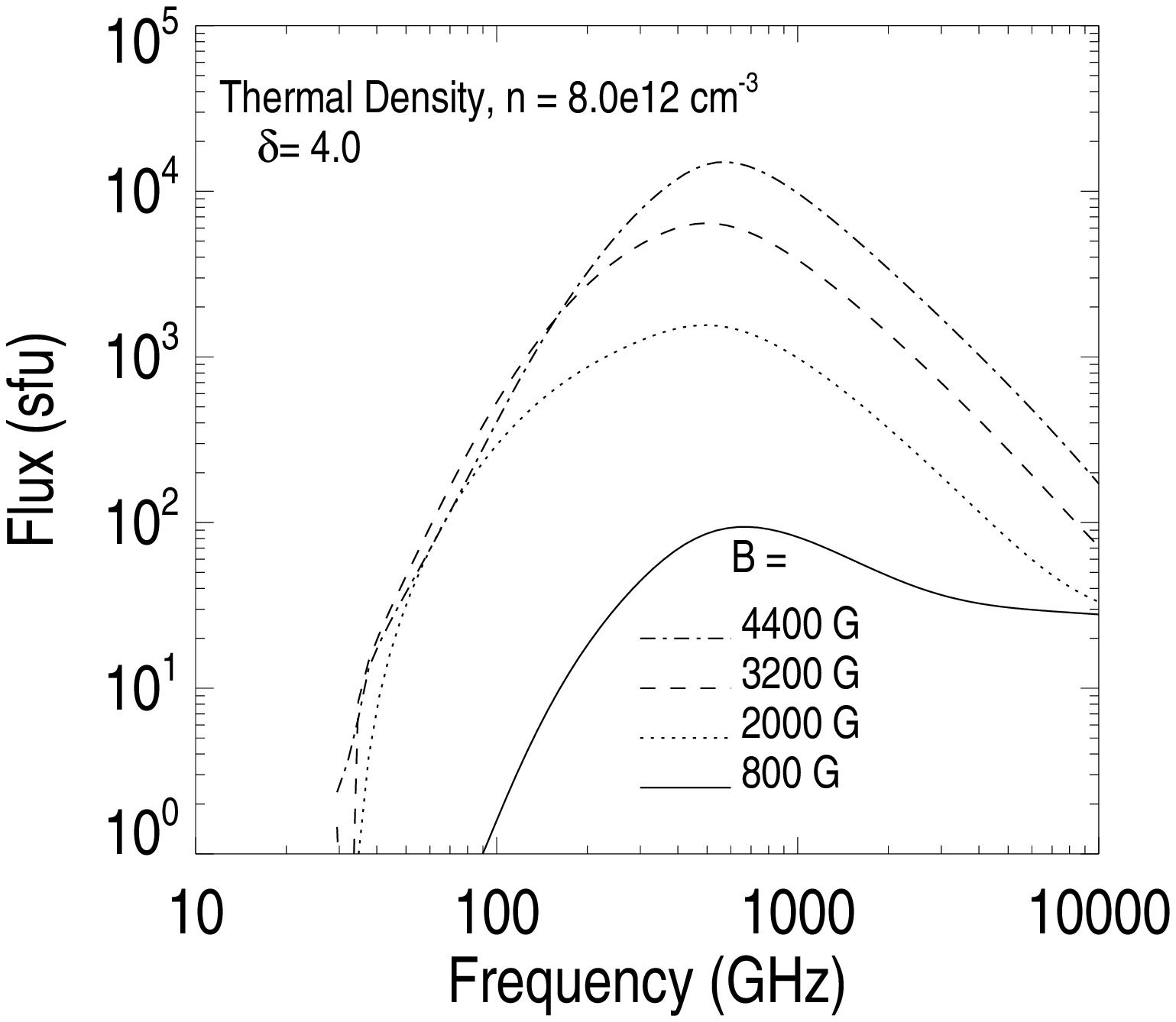}{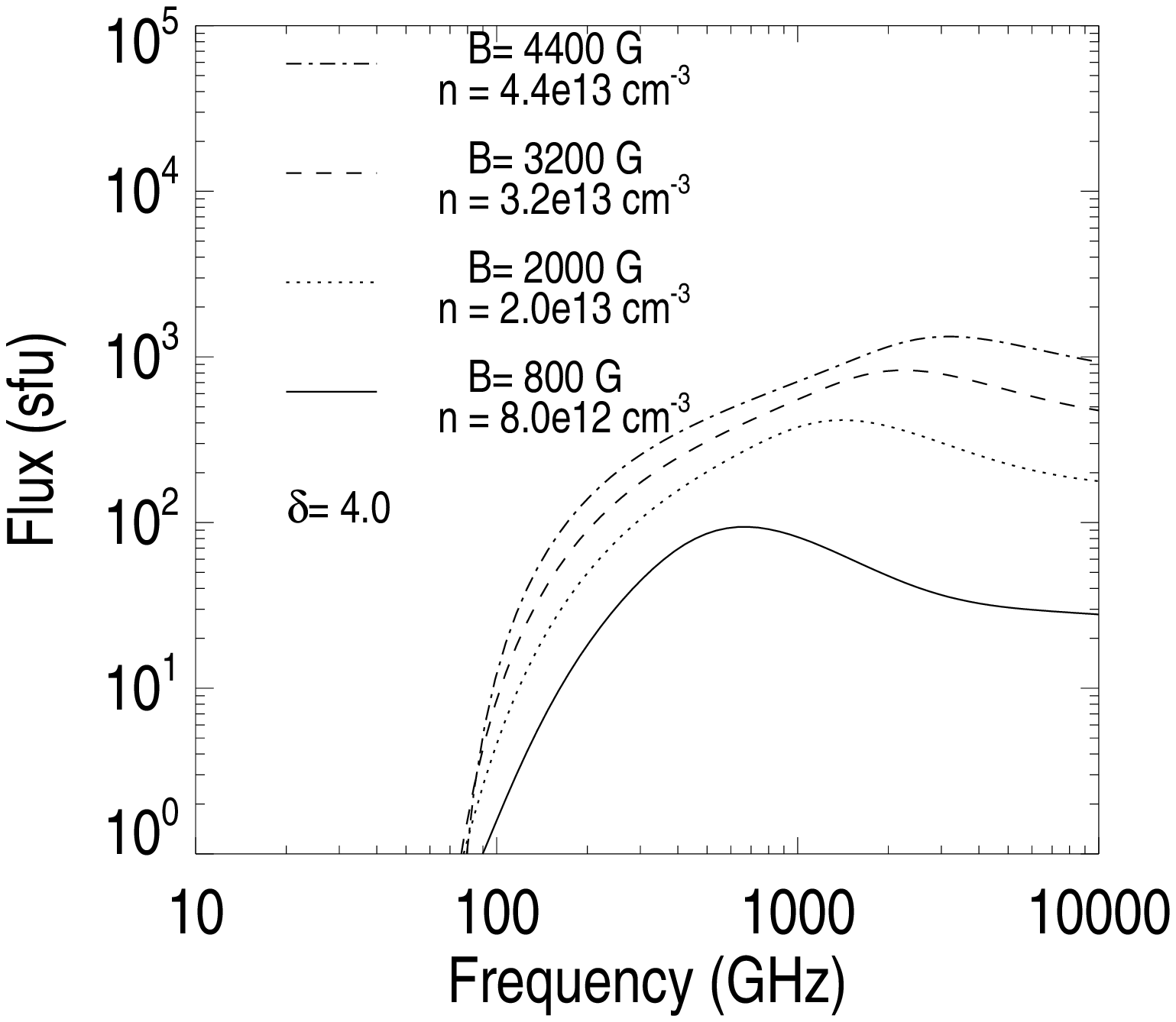}\caption{(a)Radio spectra
produced by GS plus free-free contributions from a uniform source
with a  size of $1''$ for $n_e=8\cdot10^{12}$~cm$^{-3}$ and
$B=800\dots4400$~G. (b) Razin-suppressed GS spectra with the Razin
frequency 200 GHz plus the free-free component.} \label{FIG_GS}
\end{center}
\end{figure*}

\section{Gyrosynchrotron Emission from a Compact Source}

Flare generated electrons gyrating in a magnetic field produce
strong gyrosynchrotron emission across a broad range of frequencies
\citep[e.g.][]{BBG,Nindos_etal_2008}. \citet{Silva_etal_2007}
have demonstrated that a rising optically thick GS spectrum in
the sub-THz range requires a large magnetic field in the emission source;
typically, this implies also a large plasma density. Therefore, we
include the free-free absorption and emission along with the GS
contribution in our calculations. We have investigated the
corresponding parameter space and confidently confirm the conclusion
of \citet{Silva_etal_2007} that only a very compact source
$\sim1''$ with a very strong magnetic field $B>2000$~G, in which
all flare accelerated electrons
($N_e[>50~\mbox{keV}]=5\cdot10^{35}$, $n_e[>50~\mbox{keV}]\sim10^{12}$~cm$^{-3}$)
simultaneously reside, could be consistent with the spectrum of sub-THz flare component,
Figure~\ref{FIG_GS}. The number of required electrons is typical for
a large (GOES X-class) solar flare \citep[e.g.][]{Brown_etal2007}.

Although this model is consistent with a small source size, it
requires more than extreme source parameters. In addition, this
model cannot account for large spectral indices ($>3$) of the
sub-THz component. The Razin effect in a dense plasma with a
magnetic field around 800~G was proposed to form a steeper spectrum
in the sub-THz range \citep{Silva_etal_2007}. However, as seen from
Figure~\ref{FIG_GS}, the flux density for $B=800$~G is much lower
than observed (in fact, the free-free absorption additionally
reduces the flux here). A larger magnetic field (to increase the
flux level) will require a proportionally larger thermal electron
density to keep the Razin effect in place, which will further
increase free-free absorption in the source, so the observed flux
level will not be met under conditions of strong Razin effect,
Figure~\ref{FIG_GS}(b). An increase of the source size above $2''$
with the same total number of fast electrons, magnetic field and
thermal electron density results in a spectrum totally dominated by
the free-free contribution---the model considered in the previous
section.

We note that GS emission model can easily account for the observed
temporal pulsations of the sub-THz component via, e.g., loop
oscillations at large time scales and/or fluctuations of fast
electron injection \citep{Kiplinger_etal1984,Aschwanden_etal1998} at
various time scales. In addition, good temporal correlation with
high energy hard X-rays and gamma-ray continuum will be rather
natural. Nevertheless, we feel that unrealistically extreme GS
source parameters are needed to reconcile the model with
observations of sub-THz emission from large flares. In contrast,
weaker sub-THz fluxes $F_f\sim100$~sfu observed from M-class flares
\citep{Cristiani_etal_2008} may agree with the GS emission
mechanism.

\begin{figure*}[t]
\centering
  \subfigure[]{\label{fig:edge-a}\includegraphics[width=3.3in]{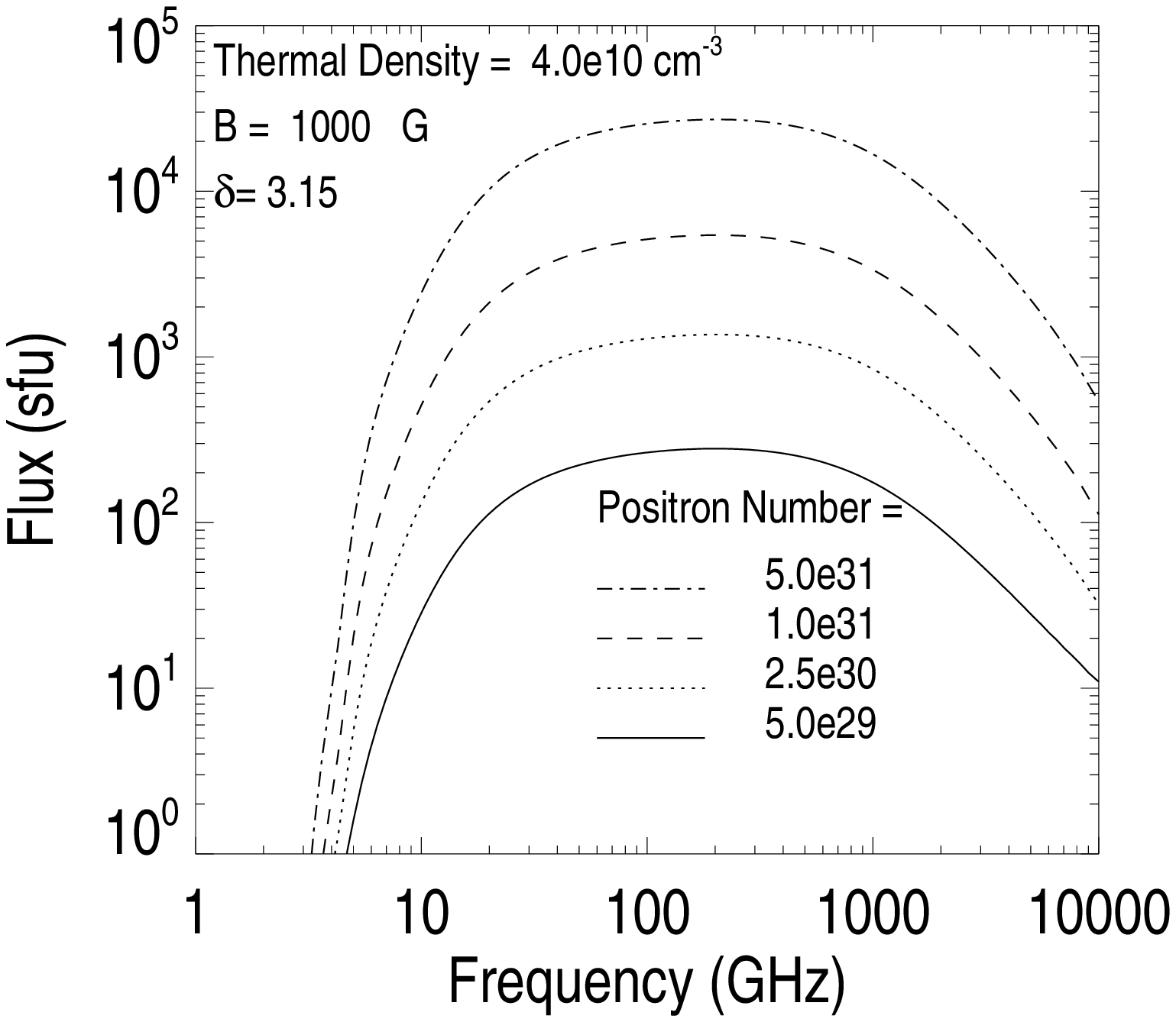}}
  \subfigure[]{\label{fig:edge-b}\includegraphics[width=3.3in]{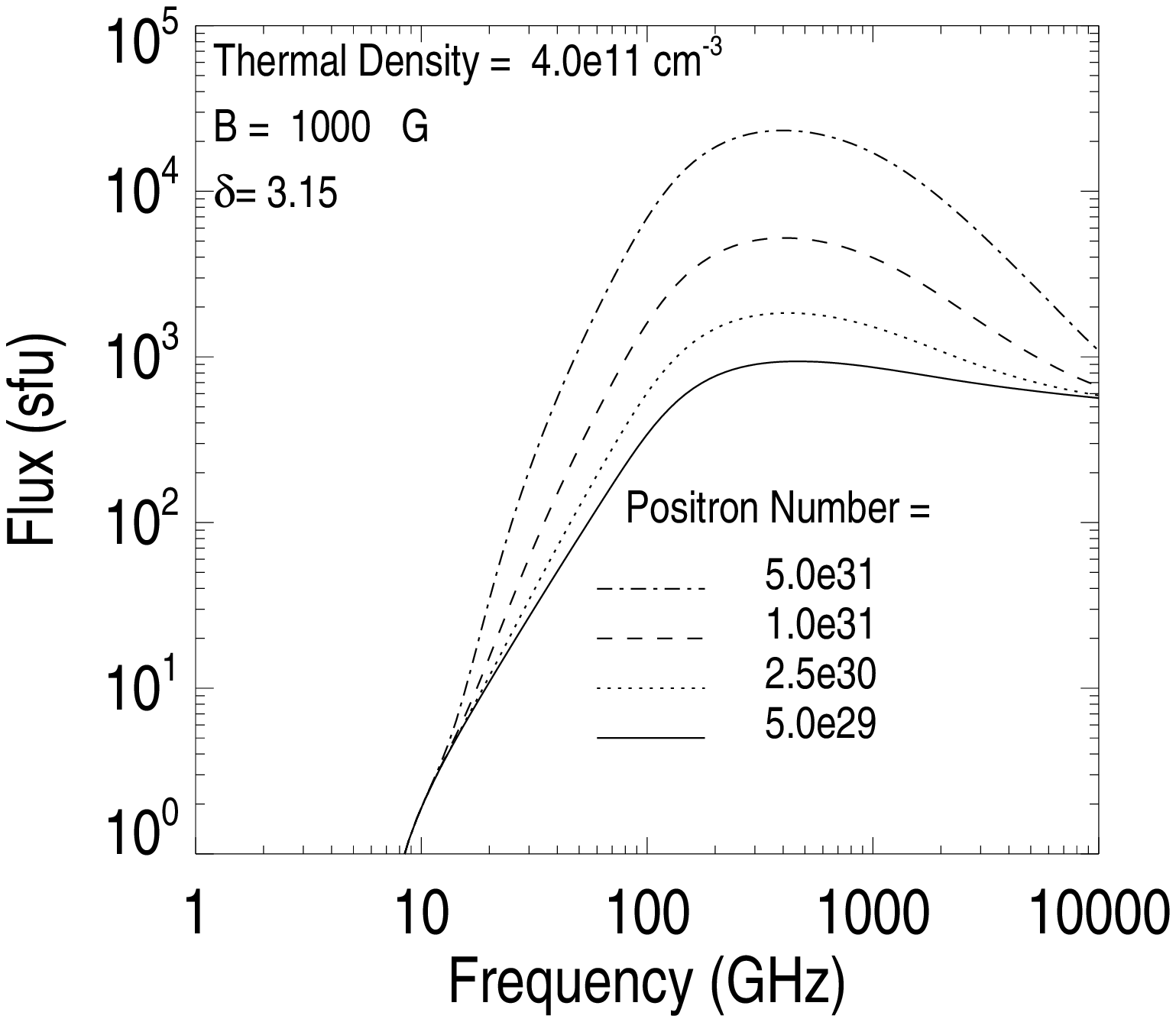}}\\
 \caption{Radio spectra produced by synchrotron radiation from relativistic
positron plus free-free contribution from a uniform cubic source
with a linear size of $20''$ for the total instantaneous positron
number $N_{e+}=5\times10^{29}\dots 5\times10^{31}$, with energy
$\gamma=20$ ($\sim$~10~MeV), magnetic field $B=1000$~G, and the
thermal electron density $n_e=4\cdot10^{10}$~cm$^{-3}$ (a) and
$n_e=4\cdot10^{11}$~cm$^{-3}$ (b), and $T_e=1$~MK. } \label{FIG_SR}
\end{figure*}

\section{Synchrotron Emission from Relativistic Positrons}

Flare-accelerated ions of tens  MeV or above colliding with thermal
ions in dense layers of the solar atmosphere trigger nuclear reactions,
with relativistic positrons being one of the products of such
interactions \citep{Lingenfelter_Ramaty_1967, Kozlovsky_etal_2002}.
These relativistic positrons are capable of producing sub-THz
synchrotron radiation with a peak frequency around
$f_{peak}{\sim}f_{Be}\gamma^2$, where $f_{Be}$ is the electron
gyrofrequency and $\gamma$ is the Lorenz-factor, as suggested by
\citet{Lingenfelter_Ramaty_1967}.

Although the synchrotron peak frequency can easily fall into the sub-THz
or THz range \citep{Trottet_etal_2008} considering typical magnetic field
($B\sim1000$~G) and Lorenz-factors $\gamma\sim20$, the low-frequency
synchrotron spectrum, $F_f{\propto}f^{1/3}$, Figure~\ref{FIG_SR}(a),
is inconsistent with the
whole range of the observed spectral index values
unless we adopt an additional absorption mechanism leading to a
sharper spectral shape. The only viable absorption mechanism is
free-free absorption, which again requires high plasma density,
$\sim10^{12}$~cm$^{-3}$,  Figure~\ref{FIG_SR}(b). Even though a
combination of the synchrotron radiation from relativistic positrons
and free-free absorption is capable of producing the required
spectral shape, the correct flux density requires more relativistic
positrons than seems to be available from nuclear interactions
even in large flares. The total number of energetic protons above 30
MeV for large solar flares as deduced from RHESSI observations is in
the range $10^{29}-10^{33}$ \citep[e.g.][]{Shih_etal2009}, which is
comparable to the instantaneous number of positrons with Lorentz
factor, $\gamma=20$  required to explain observed sub-millimeter
fluxes. The total thick-target positron yield is about
$10^{-2}-10^{-4}$ per proton of energy above 10~MeV
\citep{Kozlovsky_etal2004}, so the total number of positrons
produced in a flare should be $\lesssim10^{31}$. Since the positron
lifetime in a dense positron-production site  is likely to be less
than  duration of a flare, the instantaneous positron number of
$\sim10^{31}$ is difficult to achieve, unless we allow a
significant fraction of relativistic positrons to escape to and then
be trapped at a tenuous coronal part of the flaring loop, where the
positron lifetime is much longer.

Synchrotron radiation from relativistic positrons 
is easily consistent with small sizes if the emission region is
placed in footpoints of a flaring loop ($\lesssim7''$) or in a
moderate-size ($\lesssim20''$) coronal flaring loop; such sizes are
consistent respectively with hard X-ray measurements of footpoint
sizes \citep{Kontar_etal2008} and radio measurements of the coronal
loop sizes \citep{BBG}.

\begin{figure*}
 \begin{center}
   \subfigure[]{\label{fig:edge-a}\includegraphics[width=3.3in]{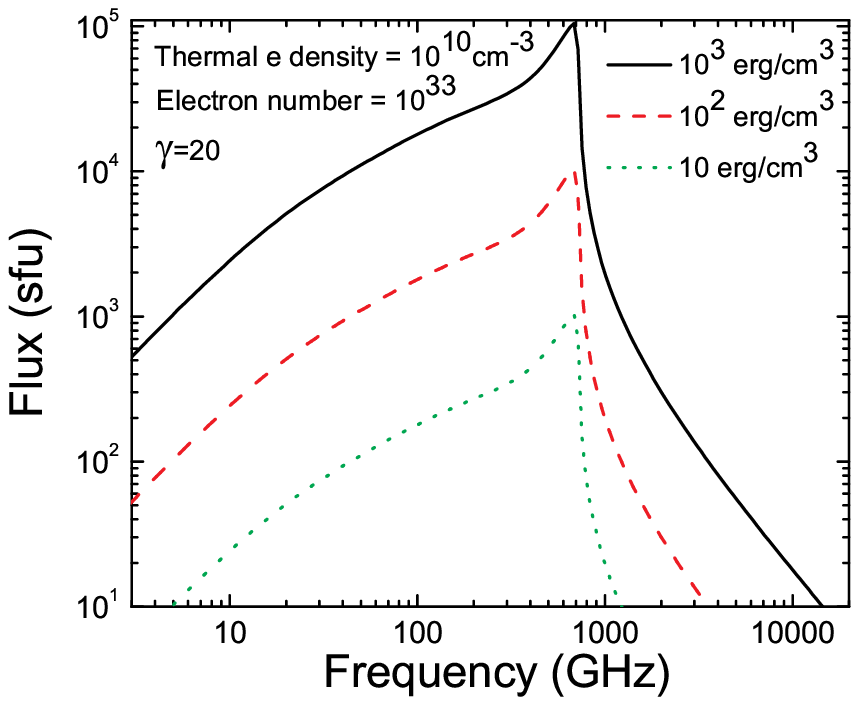}}
  \subfigure[]{\label{fig:edge-b}\includegraphics[width=3.3in]{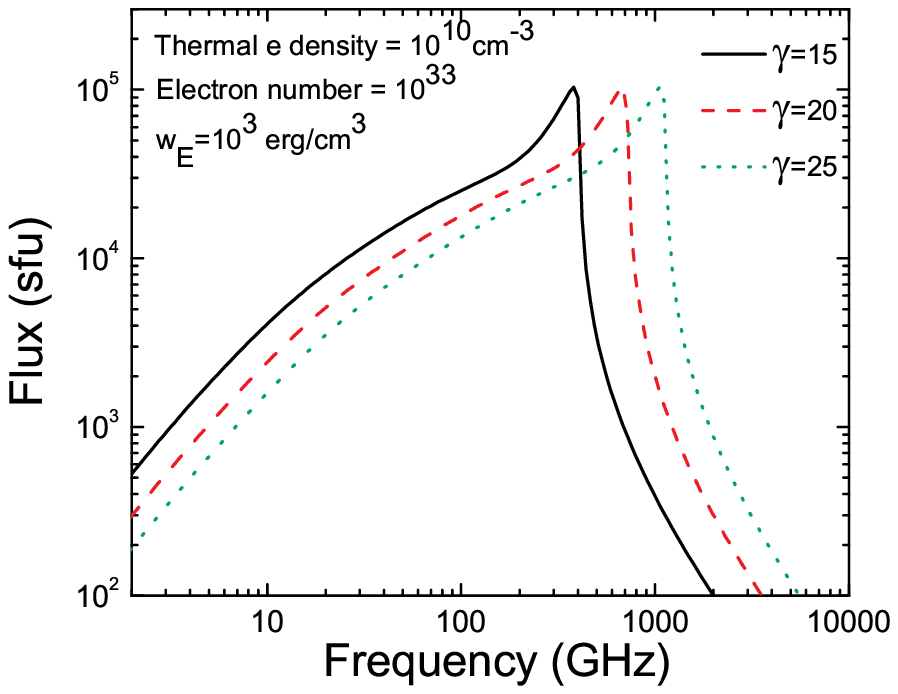}}\\
  \caption{DRL spectrum produced by relativistic electrons/positrons in long-wave Langmuir turbulence $\lambda{\gg}2{\pi}c/\omega_{pe}$. (a) Dependency on Langmuir wave energy density,
  (b) The spectra for different Lorentz factors.}
  \label{FIG_DRL}
  \end{center}
\end{figure*}

\section{Diffusive Radiation in Langmuir Waves with Large Wavelengths}

Unlike the emission processes discussed above, which require dense
plasma to produce a sufficiently fast-growing spectrum with frequency,
there is a class of wave up-scattering processes, for which a high
plasma density is not essential. 
Inverse Compton scattering of gyro-synchrotron emitted photons is
easy to estimate to be inefficient in solar flare conditions because
the energy density of electromagnetic emission is much lower than
that of the magnetic field, and $\tau_{IC}{\ll}1$. Nevertheless,
similar processes, in which the role of low-frequency photons is
played by plasma waves can be relevant.

Indeed, a flare-accelerated or decay-produced relativistic charged
particle moving through the plasma with the turbulent waves
experiences random Lorenz forces and so diffuses in space.
Accordingly, the corresponding radiative processes are called
\textit{diffusive} radiations \citep[e.g.,][]{Fl_2006a}. Generally,
the diffusive radiations do not produce a spectrum rising with
frequency \citep[e.g.,][]{Fl_2006a}. An exclusion is 
DRL for long-wavelength Langmuir waves
\citep{Fl_Topt_2007_MNRAS,Fl_Topt_2007_PhRvE}. In this case the
spectrum peaks at frequency $2f_{pe}\gamma^2$, where $f_{pe}$ is the
plasma frequency, Figure~\ref{FIG_DRL}. The spectral index can be as
large as 2. The time variability is expected due to nonlinear
dynamics of the Langmuir waves, which are known to oscillate under
nonlinear wave-wave interactions \citep[e.g.][]{KontarPecseli2002}.

\begin{figure*}[!t]
 \centering
  \subfigure[]{\label{fig:edge-a}\includegraphics[width=3.3in]{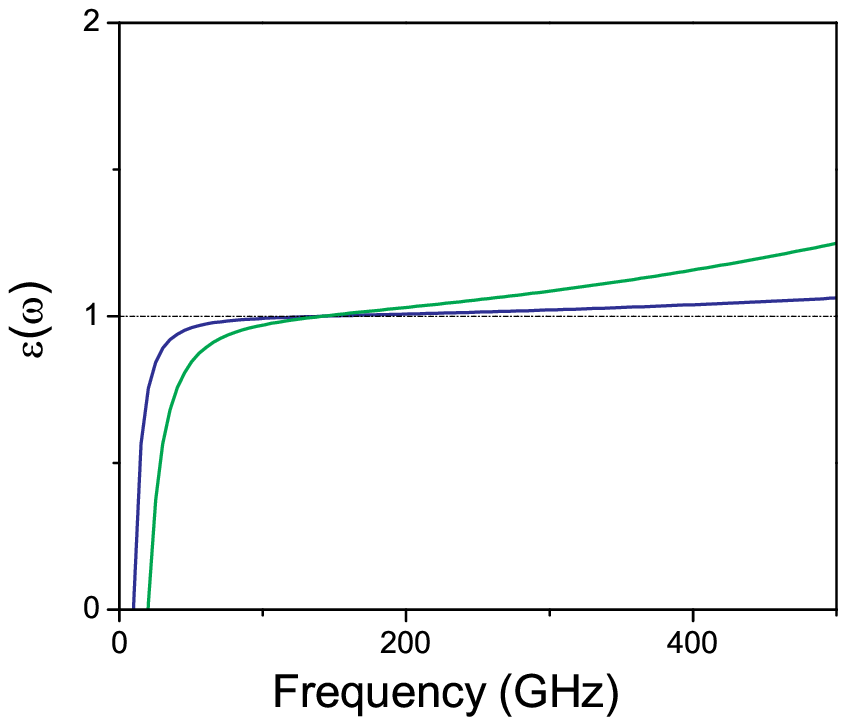}} 
  \subfigure[]{\label{fig:edge-b}\includegraphics[width=3.3in]{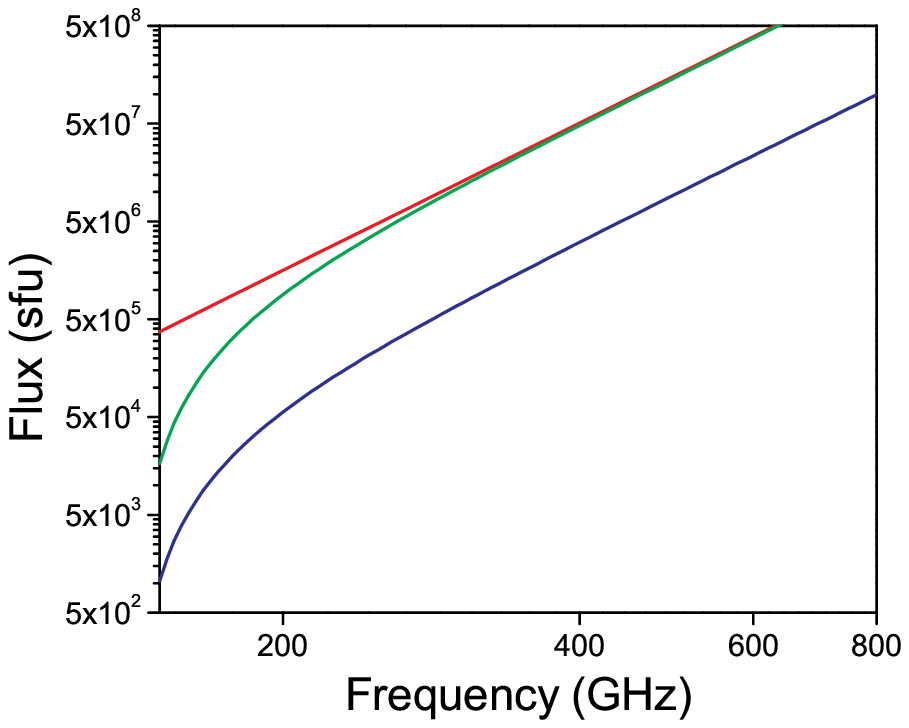}}\\
  \caption{(a) A model of plasma dielectric permittivity with molecular line contribution
  included; 
  (b) Vavilov-Cherenkov radiation produced by fast electrons with a power-law
  distribution over the velocity---blue and green curves; the red curve is for
  $\varepsilon(\omega)=1+\omega^2/\omega_0^2$, i.e.,
  without standard plasma contribution.}
 \label{FIG_VCh}
\end{figure*}

Although this emission process is attractive, it is unclear whether
there are sufficient levels of Langmuir turbulence energy density
and relativistic electron/positron numbers to produce the required
flux densities. Langmuir waves are effectively generated at shorter
wavelengths $\lambda=2{\pi}v/\omega_{pe}<2{\pi}c/\omega_{pe}$, at
the typical resonance phase velocities with the electron beam
$v=(0.2-0.6)c$ \citep[e.g.,][]{Kaplan_Tsytovich_1973, Kontar2001}.
Therefore, non-resonant generation of the plasma waves
\citep{Dieckmann_2005} or nonlinear wave-wave processes
\citep{Kaplan_Tsytovich_1973} transferring energy from small to
large  wavelength are also implied for this model.

Typically, a large level of Langmuir turbulence would result in
strong plasma radio emission due to either coalescence of the plasma
waves or scattering/decay into transverse electromagnetic waves. In
the DRL model considered here, however, most of the Langmuir
turbulence energy resides in long wavelength Langmuir waves, whose wave vectors
are small, $k\ll\omega_{pe}/c$. These long wavelength waves cannot coalescence
into transverse waves because the conservation of
momentum cannot be fulfilled in this three-wave process. Only a
minor part of the energy from Langmuir turbulence with
$k\gtrsim\omega_{pe}/c$ will be involved in the plasma radiation
production, resulting in a modest level of coherent decimeter
emission if any, which does not contradict the context radio
observations.

To round up the discussion involving the plasma waves, we note that
all nonlinear plasma processes including Langmuir waves and
\emph{non-relativistic} particles are of little interest here since
the fundamental and  harmonic plasma frequencies are well below $30$
GHz in both chromosphere and corona \citep{Aschwanden_etal2002}.


\section{Vavilov-Cherenkov Radiation from Chromospheric Layers}

Vavilov-Cherenkov radiation is produced by any charged particle
moving faster than the corresponding speed of light in a medium. In
fully ionized plasma the high-frequency dielectric permittivity
is less than unity, $\varepsilon(\omega)\lesssim1$. Therefore, the
phase velocity of electromagnetic waves
$c/\sqrt{\varepsilon(\omega)}$ is larger than the speed of light,
$c$, and the Cherenkov emission does not occur. However, the
chromospheric gas is only partly ionized; and there are numerous
atoms and molecules whose quantum transitions can make positive
contribution to the dielectric permittivity, making Cherenkov
radiation possible at certain frequency windows. Charged particle of
velocity $v$  emits Cherenkov emission in the medium, where its
dielectric permittivity $\varepsilon(\omega)$ is such that
$v>c/\sqrt{\varepsilon(\omega)}$. The dielectric permittivity of
gases is only slightly more than unity in the optical range [e.g.,
Hydrogen gas has $(\varepsilon-1)\sim2\cdot10^{-4}$], so only highly
relativistic particles with $v>(1-10^{-4})c$ can emit. The situation
is however very different in other frequency ranges, near transition
frequencies of atoms or molecules: $\varepsilon(\omega)$ goes up and
down being considerably larger than one at certain frequency
windows, which allows even sub-relativistic electrons to emit
Vavilov-Cherenkov radiation.

Each atomic or molecular quantum transition makes a contribution to
the dielectric permittivity of the form of
$\delta\varepsilon_{nm}(\omega)=\frac{4{\pi}n_{e}e^2}{m_e}\frac{S_{nm}}{(\omega_{nm}^2-\omega^2)+i\Gamma_{nm}\omega}$,
where
$S_{nm}$ is the oscillator strength of the transition, 
$\omega_{nm}$ is the transition frequency, $\Gamma_{nm}$ is the
transition decay constant. The resulting dielectric permittivity
accounting for plasma (free electrons) and molecular contributions
is
$\varepsilon(\omega)=1-\omega_{pe}^2/\omega^2+\sum\delta\varepsilon_{nm}$.
The exact spectroscopic permittivity depends on the chemical
composition of the chromosphere with the sum over all excitation
states of corresponding molecules, which is beyond our intention.
Here we perform order of magnitude estimates. We assume that there
are many atomic/molecular transitions capable of making a
contribution to the dielectric properties of gas. Then, we note that
the Cherenkov radiation spectrum rises with frequency only if the
dielectric permittivity raises with frequency. For example, the
energy level distribution between the rotational levels of
chromospheric molecules can be assumed to have a Boltzmann
distribution with the gas temperature $T$, therefore population
densities will have maxima at energy levels above the THz range
increasing to about $k_BT$. Thus, we adopt the mean molecular contribution
to the dielectric permittivity to have the  \textit{model} form
$\omega^2/\omega_0^2$, where $\omega_0$ is an unknown constant.

Assuming a power-law spectrum of fast electrons
$n_e(v)=AN_0v_0^{\beta-1}/v^{\beta}$, $v<c$,
where $N_0$ is the total number of electrons with velocity above
minimum velocity $v_0$, $\beta$ is the spectral index, and $A$ is a
dimensionless normalization constant of order of unity, the
Cherenkov emission yields flux at Earth:
\begin{equation}\label{Eq_VCh_flux}
    F_f=\frac{10^{19}}{4{\pi}R^2\au}\frac{(2\pi)^2AN_0e^2fv_0^{\beta-1}}{c^{\beta}}\times$$$$
    \left[\frac{2\varepsilon^{(\beta-2)/2}}{\beta(\beta-2)}+\frac{1}{\beta\varepsilon}-\frac{1}{\beta-2}\right]\qquad(\sfu).
\end{equation}
Figure~\ref{FIG_VCh}(a) displays the model dielectric permittivity
rising with frequency as described, while Figure~\ref{FIG_VCh}(b)
shows the corresponding Cherenkov spectra. The spectrum shape and
the flux density level allow this emission to be easily reconciled
mechanism with observations for typical numbers of accelerated
electrons. The model flux density is much larger than observed,
being $5\times10^7$~sfu at 400~GHz, which allows us to relax the
number of radiating electrons or/and add some free-free absorption
to the model. Given that the emission is from compact footpoints
$<10''$, where energetic particles interact with the chromosphere,
we conclude that the Cherenkov emission is fully consistent with the
observations of the rising sub-THz component of large solar flares.
The time variability requires corresponding fluctuations of the
electron distribution function, 
similar to sub-second variations of microwave GS or HXR radiations.

\section{Discussion and conclusions}

We have analyzed a number of emission processes, which are capable
of producing radiation at the sub-millimeter wavelengths: free-free
emission, gyrosynchrotron and synchrotron processes, diffusive
radiation, and Vavilov-Cherenkov process. Having in mind the
characteristic ranges of parameters of a solar flare region, we
calculated the spectra for each process discussed in the paper.
Although current observations at two frequencies only do not provide
us with detailed spectral shape, we estimate the range of flux
levels and spectral indices for each model and compare them with the
observations. 

It is likely that the sub-THz emission originates from more than a
single source and  more than one mechanism is involved. Free-free
emission is a plausible candidate in many cases, at least for large
sources observed by \citet{Luethi_etal_2004a,Luethi_etal_2004b}. The
free-free emission is clearly always present, so other mechanisms
build additional contributions on top of the free-free component.
Gyrosynchrotron/synchrotron emission is likely to play a role in
moderate events and also as the extension of normal microwave
bursts, which falls with frequency. The role of DRL is less clear
since the level of long-wavelength Langmuir waves is  yet unknown in
flares. Finally, the Vavilov-Cherenkov emission from compact sources
located at the chromospheric level seems to be a plausible process
to account for the rising with frequency sub-mm component of large
flares. Indeed, the dielectric permittivity of partially ionized
plasma is known to fluctuate around unity due to atomic and
molecular transition contributions. When the flare-accelerated
electrons are present, they will produce Cherenkov radiation at all
frequency windows (including IR, viz, and UV bands) where the
dielectric permittivity is (even marginally) above unity, giving
rise to a radiation spectrum raising with frequency for (mean)
dielectric permittivity increasing with frequency.

The available constraints on sub-THz source sizes, time variability,
and spectral index are not yet fully reliable particularly because
the ground-based sub-THz observations are extremely difficult to
calibrate due to strongly variable atmospheric opacity. Further
progress in understanding the physics of sub-THz emission from
flares requires observations a) with a more complete spectral
coverage at the sub-mm range (preferably well-calibrated space-based
observations) and b) polarization measurements. Nevertheless,
the sub-THz spectral window can be extremely informative, e.g., for
diagnostics of the chromospheric chemical composition if the role of
the Vavilov-Cherenkov emission is confirmed. This calls for a new
project mounting a sub-THz receivers/interferometers, combining good
sensitivity with high spatial resolution, on a space mission,
complementing on-going efforts in the microwave range.

\acknowledgments This work (GDF) was supported in part by NSF
grants ATM-0707319 and AST-0908344, and NASA grant 09-HGI09-0057,
and by the Russian Foundation
for Basic Research, grants 08-02-92228, 09-02-00226, and
09-02-00624. EPK gratefully acknowledges comments by Lyndsay Fletcher
and financial support by a STFC (UK) rolling grant, STFC Advanced Fellowship, the Leverhulme Trust
and by the European Commission through the SOLAIRE Network
(MTRN-CT-2006-035484).



\begin{thebibliography}{38}
\expandafter\ifx\csname
natexlab\endcsname\relax\def\natexlab#1{#1}\fi

\bibitem[{{Aschwanden}(2005)}]{Aschw_2005}
{Aschwanden}, M.~J. 2005, {Physics of the Solar Corona. An
Introduction with
  Problems and Solutions (2nd edition)}

\bibitem[{{Aschwanden} {et~al.}(2002){Aschwanden}, {Brown}, \&
  {Kontar}}]{Aschwanden_etal2002}
{Aschwanden}, M.~J., {Brown}, J.~C., \& {Kontar}, E.~P. 2002,
\solphys, 210,
  383

\bibitem[{{Aschwanden} {et~al.}(1998){Aschwanden}, {Kliem}, {Schwarz},
  {Kurths}, {Dennis}, \& {Schwartz}}]{Aschwanden_etal1998}
{Aschwanden}, M.~J., {Kliem}, B., {Schwarz}, U., {Kurths}, J.,
{Dennis}, B.~R.,
  \& {Schwartz}, R.~A. 1998, \apj, 505, 941

\bibitem[{{Bastian} {et~al.}(1998){Bastian}, {Benz}, \& {Gary}}]{BBG}
{Bastian}, T.~S., {Benz}, A.~O., \& {Gary}, D.~E. 1998, \araa, 36,
131

\bibitem[{{Bazylev} \& {Zhevago}(1987)}]{Bazylev_Zhevago_1987}
{Bazylev}, V.~A. \& {Zhevago}, N.~K. 1987, Moscow Izdatel Nauka

\bibitem[{{Brown} \& {Kontar}(2005)}]{BrownKontar2005}
{Brown}, J.~C. \& {Kontar}, E.~P. 2005, Advances in Space Research,
35, 1675

\bibitem[{{Brown} {et~al.}(2007){Brown}, {Kontar}, \&
  {Veronig}}]{Brown_etal2007}
{Brown}, J.~C., {Kontar}, E.~P., \& {Veronig}, A.~M. 2007, in
Lecture Notes in
  Physics, Berlin Springer Verlag, Vol. 725, Lecture Notes in Physics, Berlin
  Springer Verlag, ed. K.-L. {Klein} \& A.~L. {MacKinnon}, 65--+

\bibitem[{{Cristiani} {et~al.}(2008){Cristiani}, {Gim{\'e}nez de Castro},
  {Mandrini}, {Machado}, {Silva}, {Kaufmann}, \&
  {Rovira}}]{Cristiani_etal_2008}
{Cristiani}, G., {Gim{\'e}nez de Castro}, C.~G., {Mandrini}, C.~H.,
{Machado},
  M.~E., {Silva}, I.~D.~B.~E., {Kaufmann}, P., \& {Rovira}, M.~G. 2008, \aap,
  492, 215

\bibitem[{{Dennis} \& {Schwartz}(1989)}]{DennisSchwartz1989}
{Dennis}, B.~R. \& {Schwartz}, R.~A. 1989, \solphys, 121, 75

\bibitem[{{Dieckmann}(2005)}]{Dieckmann_2005}
{Dieckmann}, M.~E. 2005, Physical Review Letters, 94, 155001

\bibitem[{{Edwin} \& {Roberts}(1983)}]{EdwinRoberts1983}
{Edwin}, P.~M. \& {Roberts}, B. 1983, \solphys, 88, 179

\bibitem[{{Emslie} {et~al.}(2003){Emslie}, {Kontar}, {Krucker}, \&
  {Lin}}]{Emslie_etal2003}
{Emslie}, A.~G., {Kontar}, E.~P., {Krucker}, S., \& {Lin}, R.~P.
2003, \apjl,
  595, L107

\bibitem[{{Fleishman}(2006)}]{Fl_2006a}
{Fleishman}, G.~D. 2006, \apj, 638, 348

\bibitem[{{Fleishman} {et~al.}(2008){Fleishman}, {Bastian}, \&
  {Gary}}]{Fl_etal_2008}
{Fleishman}, G.~D., {Bastian}, T.~S., \& {Gary}, D.~E. 2008, \apj,
684, 1433

\bibitem[{{Fleishman} \& {Toptygin}(2007{\natexlab{a}})}]{Fl_Topt_2007_MNRAS}
{Fleishman}, G.~D. \& {Toptygin}, I.~N. 2007{\natexlab{a}}, \mnras,
381, 1473

\bibitem[{{Fleishman} \& {Toptygin}(2007{\natexlab{b}})}]{Fl_Topt_2007_PhRvE}
---. 2007{\natexlab{b}}, \pre, 76, 017401

\bibitem[{{Kaplan} \& {Tsytovich}(1973)}]{Kaplan_Tsytovich_1973}
{Kaplan}, S.~A. \& {Tsytovich}, V.~N. 1973, {Plasma astrophysics}
  (International Series of Monographs in Natural Philosophy, Oxford: Pergamon
  Press, 1973)

\bibitem[{{Kaufmann} {et~al.}(2009{\natexlab{a}}){Kaufmann}, {Gim{\'e}nez de
  Castro}, {Correia}, {Costa}, {Raulin}, \&
  {V{\'a}lio}}]{Kaufmann_etal_2009fast}
{Kaufmann}, P., {Gim{\'e}nez de Castro}, C.~G., {Correia}, E.,
{Costa},
  J.~E.~R., {Raulin}, J.-P., \& {V{\'a}lio}, A.~S. 2009{\natexlab{a}}, \apj,
  697, 420

\bibitem[{{Kaufmann} {et~al.}(2001){Kaufmann}, {Raulin}, {Correia}, {Costa},
  {de Castro}, {Silva}, {Levato}, {Rovira}, {Mandrini}, {Fern{\'a}ndez-Borda},
  \& {Bauer}}]{Kaufmann_etal2001}
{Kaufmann}, P., {Raulin}, J.-P., {Correia}, E., {Costa}, J.~E.~R.,
{de Castro},
  C.~G.~G., {Silva}, A.~V.~R., {Levato}, H., {Rovira}, M., {Mandrini}, C.,
  {Fern{\'a}ndez-Borda}, R., \& {Bauer}, O.~H. 2001, \apjl, 548, L95

\bibitem[{{Kaufmann} {et~al.}(2004){Kaufmann}, {Raulin}, {de Castro}, {Levato},
  {Gary}, {Costa}, {Marun}, {Pereyra}, {Silva}, \&
  {Correia}}]{Kaufmann_etal_2004}
{Kaufmann}, P., {Raulin}, J.-P., {de Castro}, C.~G.~G., {Levato},
H., {Gary},
  D.~E., {Costa}, J.~E.~R., {Marun}, A., {Pereyra}, P., {Silva}, A.~V.~R., \&
  {Correia}, E. 2004, \apjl, 603, L121

\bibitem[{{Kaufmann} {et~al.}(2009{\natexlab{b}}){Kaufmann}, {Trottet},
  {Gim{\'e}nez de Castro}, {Raulin}, {Krucker}, {Shih}, \&
  {Levato}}]{Kaufmann_etal_2009a}
{Kaufmann}, P., {Trottet}, G., {Gim{\'e}nez de Castro}, C.~G.,
{Raulin}, J.-P.,
  {Krucker}, S., {Shih}, A.~Y., \& {Levato}, H. 2009{\natexlab{b}}, \solphys,
  255, 131

\bibitem[{{Ka{\v{s}}parov{\'a}} {et~al.}(2005){Ka{\v s}parov{\'a}},
  {Karlick{\'y}}, {Kontar}, {Schwartz}, \& {Dennis}}]{Kasparova_etal2005}
{Ka{\v s}parov{\'a}}, J., {Karlick{\'y}}, M., {Kontar}, E.~P.,
{Schwartz},
  R.~A., \& {Dennis}, B.~R. 2005, \solphys, 232, 63

\bibitem[{{Kiplinger} {et~al.}(1984){Kiplinger}, {Dennis}, {Frost}, \&
  {Orwig}}]{Kiplinger_etal1984}
{Kiplinger}, A.~L., {Dennis}, B.~R., {Frost}, K.~J., \& {Orwig},
L.~E. 1984,
  \apjl, 287, L105

\bibitem[{{Kontar}(2001)}]{Kontar2001}
{Kontar}, E.~P. 2001, \solphys, 202, 131

\bibitem[{{Kontar} {et~al.}(2008){Kontar}, {Hannah}, \&
  {MacKinnon}}]{Kontar_etal2008}
{Kontar}, E.~P., {Hannah}, I.~G., \& {MacKinnon}, A.~L. 2008, \aap,
489, L57

\bibitem[{{Kontar} \& {P{\'e}cseli}(2002)}]{KontarPecseli2002}
{Kontar}, E.~P. \& {P{\'e}cseli}, H.~L. 2002, \pre, 65, 066408

\bibitem[{{Kozlovsky} {et~al.}(2002){Kozlovsky}, {Murphy}, \&
  {Ramaty}}]{Kozlovsky_etal_2002}
{Kozlovsky}, B., {Murphy}, R.~J., \& {Ramaty}, R. 2002, \apjs, 141,
523

\bibitem[{{Kozlovsky} {et~al.}(2004){Kozlovsky}, {Murphy}, \&
  {Share}}]{Kozlovsky_etal2004}
{Kozlovsky}, B., {Murphy}, R.~J., \& {Share}, G.~H. 2004, \apj, 604,
892

\bibitem[{{Kramer} {et~al.}(1998){Kramer}, {Degiacomi}, {Graf}, {Hills},
  {Miller}, {Schieder}, {Schneider}, {Stutzki}, \&
  {Winnewisser}}]{Kramer_etal1998}
{Kramer}, C., {Degiacomi}, C.~G., {Graf}, U.~U., {Hills}, R.~E.,
{Miller}, M.,
  {Schieder}, R., {Schneider}, N., {Stutzki}, J., \& {Winnewisser}, G.~F. 1998,
  in Society of Photo-Optical Instrumentation Engineers (SPIE) Conference
  Series, Vol. 3357, Society of Photo-Optical Instrumentation Engineers (SPIE)
  Conference Series, ed. T.~G. {Phillips}, 711--720

\bibitem[{{Lingenfelter} \& {Ramaty}(1967)}]{Lingenfelter_Ramaty_1967}
{Lingenfelter}, R.~E. \& {Ramaty}, R. 1967, \planss, 15, 1303

\bibitem[{{L{\"u}thi} {et~al.}(2004{\natexlab{a}}){L{\"u}thi}, {L{\"u}di}, \&
  {Magun}}]{Luethi_etal_2004a}
{L{\"u}thi}, T., {L{\"u}di}, A., \& {Magun}, A. 2004{\natexlab{a}},
\aap, 420,
  361

\bibitem[{{L{\"u}thi} {et~al.}(2004{\natexlab{b}}){L{\"u}thi}, {Magun}, \&
  {Miller}}]{Luethi_etal_2004b}
{L{\"u}thi}, T., {Magun}, A., \& {Miller}, M. 2004{\natexlab{b}},
\aap, 415,
  1123

\bibitem[{{Nakariakov} \& {Verwichte}(2005)}]{NakariakovVerwichte2005}
{Nakariakov}, V.~M. \& {Verwichte}, E. 2005, Living Reviews in Solar
Physics,
  2, 3

\bibitem[{{Nindos} {et~al.}(2008){Nindos}, {Aurass}, {Klein}, \&
  {Trottet}}]{Nindos_etal_2008}
{Nindos}, A., {Aurass}, H., {Klein}, K.-L., \& {Trottet}, G. 2008,
\solphys,
  253, 3

\bibitem[{{Shih} {et~al.}(2009){Shih}, {Lin}, \& {Smith}}]{Shih_etal2009}
{Shih}, A.~Y., {Lin}, R.~P., \& {Smith}, D.~M. 2009, \apjl, 698,
L152

\bibitem[{{Silva} {et~al.}(2007){Silva}, {Share}, {Murphy}, {Costa}, {de
  Castro}, {Raulin}, \& {Kaufmann}}]{Silva_etal_2007}
{Silva}, A.~V.~R., {Share}, G.~H., {Murphy}, R.~J., {Costa},
J.~E.~R., {de
  Castro}, C.~G.~G., {Raulin}, J.-P., \& {Kaufmann}, P. 2007, \solphys, 245,
  311

\bibitem[{{Trottet} {et~al.}(2008){Trottet}, {Krucker}, {L{\"u}thi}, \&
  {Magun}}]{Trottet_etal_2008}
{Trottet}, G., {Krucker}, S., {L{\"u}thi}, T., \& {Magun}, A. 2008,
\apj, 678,
  509

\bibitem[{{Trottet} {et~al.}(2002){Trottet}, {Raulin}, {Kaufmann},
  {Siarkowski}, {Klein}, \& {Gary}}]{Trottet_etal_2002}
{Trottet}, G., {Raulin}, J.-P., {Kaufmann}, P., {Siarkowski}, M.,
{Klein},
  K.-L., \& {Gary}, D.~E. 2002, \aap, 381, 694

\end{thebibliography}

\end{document}